# Patterns, transitions and the role of leaders in the collective dynamics of a simple robotic flock

**Norbert Tarcai[1], Csaba Virágh[1], Dániel Ábel[1], Máté Nagy[1], Péter L. Várkonyi[2], Gábor Vásárhelyi[1] and Tamás Vicsek[1,3]**

[1]Eötvös Loránd University, Faculty of Science, Department of Biological Physics, Hungary
[2]Budapest University of Technology and Economics, Department of Mechanics Materials and Structures, Hungary
[3]Statistical and Biological Physics Research Group of the Hungarian Academy of Sciences, Hungary

(tarcai, viraghcs, abeld, nagymate, vasarhelyi, vicsek)@hal.elte.hu, vpeter@mit.bme.hu

**Abstract.** We have developed an experimental setup of very simple self-propelled robots to observe collective motion emerging as a result of inelastic collisions only. A circular pool and commercial RC boats were the basis of our first setup, where we demonstrated that jamming, clustering, disordered and ordered motion are all present in such a simple experiment and showed that the noise level has a fundamental role in the generation of collective dynamics. Critical noise ranges and the transition characteristics between the different collective patterns were also examined. In our second experiment we used a real-time tracking system and a few steerable model boats to introduce intelligent leaders into the flock. We demonstrated that even a very small portion of guiding members can determine group direction and enhance ordering through inelastic collisions. We also showed that noise can facilitate and speed up ordering with leaders. Our work was extended with an agent-based simulation model, too, and high similarity between real and simulation results were observed. The simulation results show clear statistical evidence of three states and negative correlation between density and ordered motion due to the onset of jamming. Our experiments confirm the different theoretical studies and simulation results in the literature about collision-based, noise-dependent and leader-driven self-propelled particle systems.

## 1. Introduction

Collective motion is one of the most spectacular phenomena in nature that appears universally in many life forms regardless of their level of intelligence or taxonomical classification [1]. Herds of cows, flock of birds [2], schools of fish [3], swarms of locusts [4] or migrating patterns of bacteria or cells [5] all fascinate biologists, physicists and non-scientists, too. Being such a widespread phenomenon, it has been hypothesised that the underlying principles that generate these complex non-equilibrium ordered states are quite simple and non-specific. To prove this, several theoretical models has been constructed successfully over the last two decades to describe different aspects of ordering, but very few experiments could verify these results so far. In this article we try to fill this gap with a very simple experimental setup of a circular pool with dozens of radio-controlled boats inside. We demonstrate that inelastic collisions with proper boundary conditions are sufficient for generating long term order and that a few individuals having a fixed orientation preference can lead a whole group of particles through solely these collisions.

*1.1. Theoretical models of flocking*
The first successful statistical physics type model describing flocking behaviour was proposed by Vicsek et al. [6]. This model consisted of self-propelled particles (SPP) with constant speed and variable direction, obeying nearest-neighbour averaging rules with added random noise in the angle updates. It was found in the simulations of this model (which we shall refer to, following Huepe and Aldana [7], as the SVM) that the boundary conditions and the noise level have a fundamental role in establishing long term order. The model became the general basis of many other extended theoretical works on collective motion. A decent number of papers have been published concerning the still ongoing debate about the order of the phase transition in the SVM model. Grégoire et al. [8] introduced the so-called vectorial noise model and demonstrated that its ordering phase transition appears to be sudden, first-order type. Aldana et al. ([9], [10]) also investigated the order of the phase transition and concluded that it depends on the intrinsic or extrinsic character of the noise in the velocity updates. Gregoire et al. [11] added adhesion to the model to force ordering in open boundary conditions. Chaté et al. [12] examined a bipolar version of the original model and observed giant density fluctuations in their simulations.

The above mentioned and many other versions of the SVM all add complexity to the original model to gain more insight into the details of the collective states. In contrast, Grossman et al. [13] made attempts to simplify the original concept and changed the neighbour-averaging velocity alignment ability of the particles to the passive communication mechanism of inelastic collisions. Quite surprisingly, they also found large-scale vortices and coherent group migration in their simulation model.

*1.2. Artificial flocking experiments with inherently self-propelled units*
Flocking is principally a natural phenomenon that is based on the local communication of living individuals, therefore human-coordinated experiments with finely tuneable parameters are quite rare.

The most general approach for creating such a setup with many identical non-living particles involves apolar or polar particles moving quasi-ballistically on a vibrated stage and possibly aligning themselves through collisions. Blair et al. [14] vibrated granular rods and observed vortex patterns in a circular pool. They could also generate net motion of the apolar rods in an annular pool due to the boundary conditions. Narayan et al. [15] used elongated apolar particles that ordered themselves in an active nematic liquid crystalline phase, showing long-lived giant number fluctuations, but the apolar fashion in the circular pool could not give rise to true net collective motion. Kudrolli et al. [16] shaked polar granular rods but the most significant collective behaviour they observed was the ordered jamming phase. Recently, Deseigne et al. [17] used vibrated disks with built-in polar weight asymmetry and could observe large scale collective motion and giant number fluctuations for short periods of time. They could even adjust the amplitude of the vibration and thus the persistence length of the particles (i.e. the noise level) to some extent, but only in the critical region between the ordered and disordered state and not deep in the ordered phase. They used very special boundary conditions that do not seem to support long-term collective motion in a bounded region.

Another generally used approach for creating an artificial flock involves intelligent robots that sense their environment usually through proximity sensors and communicate with each other wirelessly. Swarms of up to 100 self-organized units have been reached so far in 2D [18] and there are already a limited number of approaches for 3D flocks as well ([19],[20]). However, in these systems alignment and flocking is the consequence of intelligent behaviour, not passive dynamics.

In our non-living flocking experiments we manoeuvre between the two above mentioned approaches. In one hand, we use simple radio-controlled swimming robots that are truly self-propelled, on the other hand they have no intelligence at all for swarming. Self-organizing in our case is the outcome of inelastic collisions in the annular swimming pool, like in the rod-shaking experiments. However, one of the advantages of our method over the non-robotic approaches is that we can easily tune the noise level of the system in a wide range. Furthermore, our system can be extended easily with intelligent individuals, which guide the flock.

*1.3. The role of leaders in group behaviour*

The role of leaders in collective decision making has become a hot topic recently, based on widespread quantitative observations of group behaviour in biological systems (see e.g. [21] or [22]). By now, simple statistical physics models and active control experiments involving living and non-living units also appeared. Couzin et al. [23] proposed a good general leadership model and showed that a minority of informed individuals in a group can affect the behaviour of the whole group, and that for larger groups, a smaller proportion of informed individuals is enough for guiding the flock. Nagy et al. [24] investigated homing pigeon flocks and found a sophisticated leadership system: a well defined multi-level structure in which all individuals change their direction following the ones immediately above them in the hierarchy. Freeman and Biro [25] also found complex group structure among homing pigeons, however, they also demonstrated that simple interactions are enough for group navigation. Yates et al. [4] observed locusts in a ringed-shaped arena – similar in shape to the pool of our experiments – and found that locusts increase the randomness of their movements for a short period of time when changing flock direction. Although there are most probably no leaders among locusts, it might be useful for leaders of other flocks to know that collective decision making (in the sense of turning an ordered flock into a new direction) is possibly the most efficient at the critical noise level.

There are also some successful attempts to control an animal group through artificial individuals. Sumpter et al. [3] used guided fish replicas and came to the conclusion that fish can make accurate decisions without the need for complicated comparison of the information they possess. They showed that fish need no hierarchy but large group size to make good decisions. As group size increases, the many simple individual decisions result in a higher accuracy consensus of e.g. a new direction to turn into. Vaughan et al. [26] used a robotic sheepdog to guide ducks successfully to a predefined point with real time observation of the flock position and active control of the artificial duck. Correll et al. [27] electrically shocked selected cows in a herd to make them move towards the herd centre because of the induced stress, thereby moving the whole group towards a predefined position. Halloy et al. [28] successfully manipulated the shelter-selection behaviour of cockroaches through socially integrated robot-cockroaches.

Guiding a robotic flock with informed individuals is a very recent issue, mostly because robotic flocks themselves are very rare, too. Most experiments dealing with swarming robots use some kind of self organizing intelligence to keep individuals together ([29], [30], [31], [32]). However, this method also works against the goal of a possible leader whose task is to "convince" the others about its own decision. To our knowledge the only truly successful attempt to guide a robotic flock through informed individuals is provided by Celikkanat and Sahin [33].

Without the intention to provide a general control scheme for leader-driven robot swarms, we show that even the simplest collision-based information exchange in such a system can lead to a proper collective decision making mechanism under specific conditions. In our experiments we make use of the simplest possible leadership model consisting of simple self-propelled particles and a few leaders with well defined directional preferences. We show experimentally that this simple setup is sufficient to form a guided flock. We also investigate the role of noise level in the efficiency of guiding.

**2. Experimental setup**

The experimental setup consists of up to 30 commercial radio-controlled boats, a plastic pool, an industrial camera, a recording and tracking software and a PC that runs the software and controls the boats (figure 1a). For the experiments with actively driven leader boats the setup has been complemented by closed-loop real time tracking, which is used to drive these boats (figure 1b).

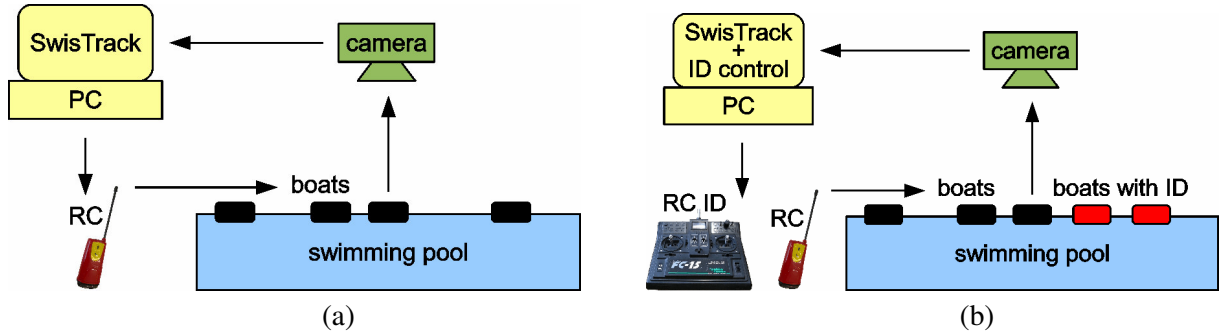

**Figure 1.** Schematic view of the experimental setup: (a) main experiment with passively controlled boats; (b) enhanced setup including leader boats with real-time closed-loop drive.

*2.1. Units – the boats*
Silverlit RC Hovercraft type commercial boats were used in our experiments (figure 2a). These simple, slow and small devices are capable of doing three things: a) go forward (*fw* phase); b) go backward and turn in a random direction (*bw* phase) due to the rudder now being in the front and making manoeuvring unstable; c) just float without propelling force (*nop* phase). In the experiments we used a sequence of *fw-nop-bw-nop* phases to avoid too high speed of *fw*-only mode and to be able to introduce a random noise into the dynamics through the *bw* phase. This way, the noise level could be adjusted precisely with the length of the *bw* phase.

It is important to emphasize that the boats had no local control; they were completely passive, individual units, "communicating" with each other only through collisions.

To aid the tracking algorithm with enhanced contrast, the boats were painted black and a retro-reflexive elongated white marker was placed on them in the middle. Since the boats had only one propeller, the rudders also had to be adjusted to compensate for the torque effect thus forcing the boats to go straight. The boats were powered with two AA type batteries that were recharged after every 40 minutes of experiments to eliminate slower speed runs due to the otherwise possible low battery state.

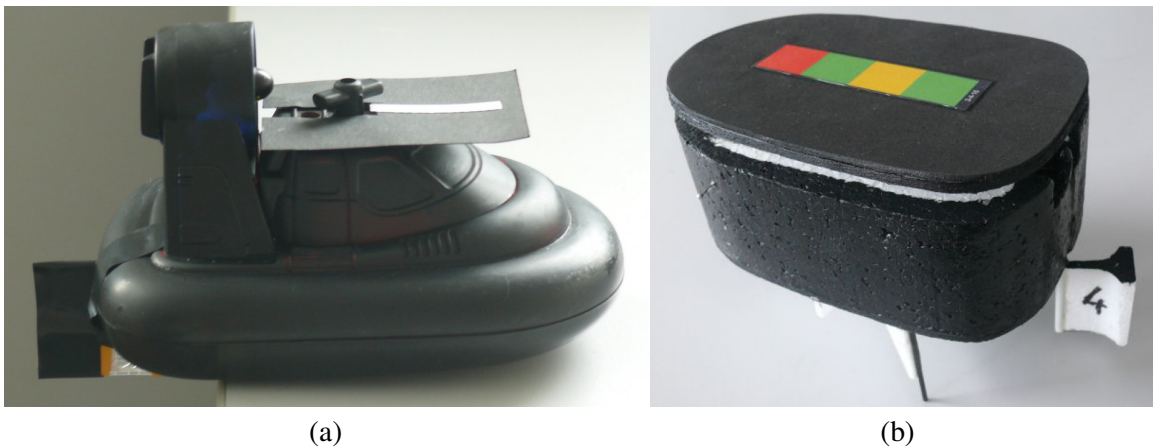

**Figure 2.** The simple robotic units. (a) The commercial boats used in the experiments were modified: superfluous parts removed, rudder fixed, painted black, retroreflective stripe attached; (b) Custom made steerable leader boats with similar properties as the normal boats, but with coloured unique barcode IDs on top.

*2.2. Boundary condition – the pool*
Computer simulations of collective motion usually apply open space or periodic boundary conditions to provide enough time and space for the particles to order. However, in real experiments only highly bounded regions can be inspected practically, where long-term order is usually harder to achieve partly due to the lack of time in open space needed for ordering and partly because – as a consequence of the boundary – ordered moving states are overtaken by jamming. Deseigne et al. [17] used a flower-like 2D space to turn particles hitting the outer bound back into the open space. This boundary shape prevents the onset of jamming, but gives no chance for long-term ordered motion. In our experiments we used an annular space to provide a partly jamming-free periodic bound in two dimensions, resulting in a quasi-one-dimensional setup with much narrower radial space than circumferential.

It is interesting to note, that in quasi-1-dimensional non-equilibrium systems, jamming is usually a consequence of particles moving in a narrow space with opposite direction [34], however, in our system this rarely happens, because a) the width of the annulus is much greater than the size of the units; b) the units are floating without friction, therefore, zero-impulse jamming clusters in free space are not stable. In our experiments, jamming is always present at the outer wall and it is fundamentally caused by the stable driving forces perpendicularly to the outer wall. So contrary to the previous case, jamming in our system can be even suppressed by changing the pool's shape from circular to annular, thereby reducing motion in the radial direction.

In our setup a simple circular swimming pool served as the outer bound for the boats and a smaller, circular obstacle was placed in the middle to give the inner bound. The resulting total space was 500 times the area occupied by a single particle, with a width-length ratio of around 1:10.

*2.3. Coordinate system*
Since we use a circular pool, the position of the boats is always referenced in polar coordinates fixed to the centre of the pool. To define the orientations of the particles, local-angle polar (LAP) coordinates are used with the angle between the local radial direction and the orientation of the boat in the $(-\pi,\pi]$ range. The schematic view of the pool with the boats and the definition of the coordinate system can be seen in figure 3.

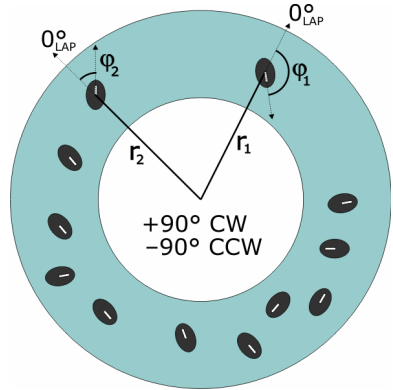

**Figure 3.** Schematic view of the pool with the boats. The local-angle polar (LAP) coordinate system defines the representation of the positions and orientations of the particles: $r$ is defined as the distance from the center of the pool, $\varphi$ is the angle between the local radial direction and the orientation of the boat.

*2.4. Camera*
A Basler scA1300 - 32gc gigE type industrial camera was placed above the centre of the pool with a LED-line around it to give light to the retroreflective markers on the boats. The 800×800 pixel images resulted in 2.5mm/pixel resolution. The image of the camera was calibrated with a point grid to compensate for the barrel-distortion of the lens and to give real-space coordinates for each pixel. Images were captured at 10 frame/s (FPS).

*2.5. PC with tracking software and boat control*

SwisTrack [35], an open-source multi-target tracking software was used to determine time, position and orientation information about each boat in each video frame. Videos and particle data were saved and analyzed offline.

The phase control of the boats was computer-driven with a fixed repeated sequence in each experiment. The control signal was sent through a parallel port and relays connected to the common radio controller of all boats. The actual phase control information was also saved along with each video frame to aid later analysis.

Each measurement was conducted with a constant control sequence starting from a random initial arrangement of the boats. Measurement length varied between 8 and 16 minutes.

*2.6. Extended setup with leader boats*

In the experiments with leaders up to four "intelligent" boats were controlled individually to align their orientation in each frame to a pre-defined angle (tangential clockwise (CW) or counter-clockwise (CCW) orientation in the pool). Thus we could determine experimentally whether a few individuals with determined orientation could 1) define the direction of the others by driving them into a collective state; 2) increase the order compared to non leader driven experiments.

The leader boats were constructed from commercial modelling components and were designed to have similar weight, size, shape, speed and power as the normal boats. They were separately controlled through an 8-channel analogue radio controller (Robbe-Futaba FC-15) that was connected to the PC's parallel port . All leader boats had an individual steering signal and a common adjustable speed signal that was switched on in a time-synchronized manner with the *fw* phase of the control signal of the normal boats.

Individual ID recognition was performed with a custom component for SwisTrack and unique coloured barcodes painted on the leader boats (figure 2b). The steering program's driving function mapping the measured actual orientation of the boats to the controlled orientation of the servo-rudder was linear with a constant cut-off above 30 degrees of misalignment. This steering algorithm resulted in a smooth circular motion in open space with high stability to perturbations and great manoeuvrability in jamming situations.

The leader experiments were either simple 8 minute runs (CW or CCW) or a continuous flow of 4 shorter 4 minute runs (CCW-CW-CCW-CW), with CW and CCW denoting the fixed preset orientation of the leader boats.

Parameters of the experimental setup are summarized in table 1.

**Table 1.** Experiment parameters.

| Parameter | Value | Unit |
|---|---|---|
| Outer diameter of the annular pool | 180 | cm |
| Inner diameter of the annular pool | 97 | cm |
| Size of boats | 10×15 | cm |
| Total space | 500 | boats |
| Total number of boats (including leaders) (n) | 27 | boats |
| Number of leader boats (k) | 0–4 | boats |
| RC control phase sequence [*fw nop bw nop*][a] | [2 1 *x* 1], where $x \in [0.4, 2.8]$ | s |
| Open space speed of boats in *fw* phase | 15 ± 5 | cm/s |
| Open space angular velocity of boats in *bw* phase | ±1 ± 0.2 | rad/s |
| Image size | 800×800 | pixel |
| Image resolution | 2.5 | mm/pixel |

| Parameter | Value | Unit |
|---|---|---|
| Recording frame rate | 10 | FPS |
| Measured individual parameters in each frame | position, orientation | |
| Measured common parameters in each frame | time, control signal, video frame | |
| Length of normal measurements | 8–16 | min |
| Length of leader measurements | 1×8 (CW or CCW) | min |
| | 4×4 (CCW-CW-CCW-CW) | |

[a] noise level in the experiments is controlled through the length of the rotational *bw* phase

## 3. Methods of analysis

Besides the detailed visual observation of the videos of the experiments we analyzed the raw data of positions, orientations, control signals and timestamps statistically, offline. We derived two main types of results from the measurement data:

1. averaged order parameters in every frame from the positions and orientations of the boats to analyze the collective patterns and the transitions between them in time,
2. probability distribution function (PDF) of the order parameters between measurements of different noise levels to analyze data in the noise space.

*3.1. Data pre-processing*

The object tracking algorithm had a 99.7% average recognition rate, so as a first step, missing single data points had to be filled using linear interpolation from neighbouring frames. After this, velocities were calculated from time, position and orientation data. The detected $[0,\pi]$ orientation of the apolar stripes also had to be projected into $(-\pi,\pi]$ using calculated velocities and the global assumption that in the *fw* phase the boats go mostly forward. Finally, a transformation from Cartesian to LAP coordinates had to be applied to be able to compare individual orientations and velocities in the annular space. From now on, if no other indications apply, we always refer to LAP coordinates without notice.

*3.2. Order parameters*

Using averaged boat orientations and velocities, several order parameters were calculated for every frame. The magnitude of the average orientation vector $\Psi_o(t)$ (1) and the orientation correlation $C_o(t)$ (2) are both normalized, continuous order parameters with a value of 0 indicating total disorder and 1 indicating full order:

$$\Psi_o(t) = \frac{1}{N}\left|\sum_{i=1}^{N} \mathbf{o}_i(t)\right|, \qquad (1)$$

$$C_o(t) = \frac{2}{N(N-1)} \sum_{\substack{i,j=1 \\ i \neq j}}^{N} \left(\mathbf{o}_i(t) \cdot \mathbf{o}_j(t)\right), \qquad (2)$$

where *N* is the number of boats and $\mathbf{o}_i(t)$ is the unit-length orientation vector of the *i*th boat.

When the orientations of the boats are correlated, the angle of the average orientation vector

$$\varphi_o(t) = \arg\left(\frac{1}{N}\sum_{i=1}^{N} \mathbf{o}_i(t)\right) \qquad (3)$$

determines whether the ordered state (figure 4) is jammed towards the outer boundary of the pool ($\varphi_o = 0$) or the boats are moving in flocks in the CW ($\varphi_o = \pi/2$) or CCW ($\varphi_o = -\pi/2$) directions.

Orientation-based order parameters are not very sensitive to the changes between the *fw-nop-bw-nop* phases in the control sequence, but the *bw* phase always decreases the level of order a bit, thereby making the order parameters in time jagged (which also shows that the *bw* phase can be used as adjustable random noise). Nevertheless, they alone give no information about the dynamics of the collective states, therefore velocity-based order parameters were also calculated the same way as the orientation-based ones.

The magnitude of the average velocity $\Psi_v(t)$ (4) and the velocity correlation $C_v$ (5) are not normalized order parameters:

$$\Psi_v(t) = \frac{1}{N}\left|\sum_{i=1}^{N} \mathbf{v}_i(t)\right|, \quad (4)$$

$$C_v(t) = \frac{2}{N(N-1)} \sum_{\substack{i,j=1 \\ i \neq j}}^{N} \left(\mathbf{v}_i(t) \cdot \mathbf{v}_j(t)\right), \quad (5)$$

where $\mathbf{v}_i(t)$ is the velocity of the *i*th boat. The value of $\Psi_v(t)$ and $C_v(t)$ highly depends on the *fw-nop-bw-nop* changes since in the *bw* phase the velocities are much smaller, the boats go backward slowly and rotate. Therefore, these order parameters have large amplitude undulations with the fundamental frequency of the control sequence. Local large velocity correlations, however, always indicate a collectively moving state, but ordered jamming can not be detected well with these parameters, since in the jammed state the velocities are small and only the orientations are correlated.

Finally, the angle of the average velocity $\varphi_v(t)$ (6) is defined analogously to $\varphi_o(t)$:

$$\varphi_v(t) = \arg\left(\frac{1}{N}\sum_{i=1}^{N} \mathbf{v}_i(t)\right) \quad (6)$$

### 4. Experimental results
From the statistics of approximately 50 measurements with 27 boats and all together 12 hours of active group motion we could clearly identify all previously simulated states of collective motion and could observe many details regarding the transitions between them. In the statistical analysis we only used data from the second half of the *fw* control phase and the first half of the consecutive *nop* phase. This was the range where the boats moved more or less ballistically due to propelling forces and inertia.

*4.1. Observed patterns of collective motion*
The different states of collective motions that could be observed during our experiments are illustrated in figure 4.

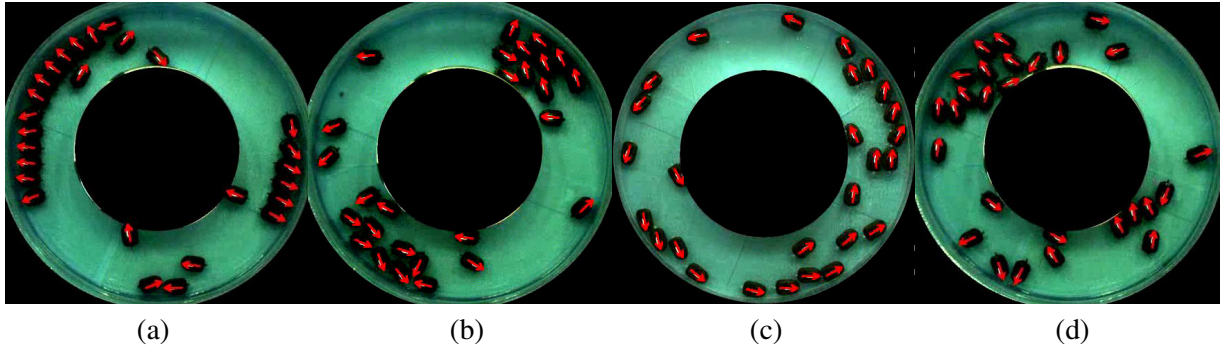

(a)            (b)            (c)            (d)

**Figure 4.** Snapshots of the camera above the pool with the arrows attached to the automatically tracked particles (pointing forward). The following patterns could be observed in the experiments: (a) jamming; (b) clustering (as an instantaneous phenomenon); (c) ordered collective motion in CW or CCW direction; (d) disordered motion.

The main characteristics of the observed states are summarized in table 2.

**Table 2.** Characteristics of observed collective patterns.

| Name of pattern/state | Typical noise level | Regularity | Stability |
|---|---|---|---|
| **Jamming** | low | sometimes; more at very low noise level | stable |
| **Clustering** | middle | rare, only momentarily during transitions between more stable states | very unstable |
| **Ordered motion** | low-middle | Regular | low noise: stable; high noise: unstable |
| **Disordered motion** | high | Regular | high noise: stable; low noise: unstable |

*4.2. State transitions in time*

During the experiments we observed transitions between the jamming-ordered, jamming-disordered and ordered-disordered states[1]. In general, velocity-based order parameters have more abrupt changes during state transitions; they can change their value between extremities during one single control phase cycle (a few seconds). The orientation-based order parameters change more slowly but they signal the forthcoming state transitions earlier. This is in correspondence with the visual observations that e.g. a jammed flock can be ordered next to the boundary without notable translational motion, then very suddenly switch state into a coherent moving flock. Order parameters of a typical case at mid noise level ($bw = 1.4$ s) can be seen in figure 5.

---

[1] These transitions are almost certainly the analogues of the *phase transitions* of similar non-equilibrium systems of very high particle number, but with this number of experimental boats we can only call them *state transitions*.

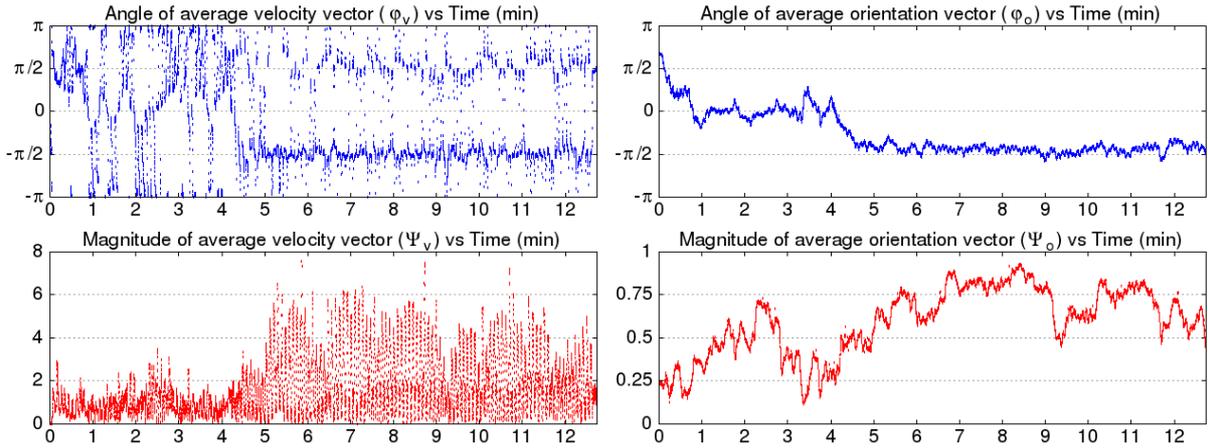

**Figure 5.** Order parameters during an experiment at *bw* = 1.4 s noise level (mid range). Starting from random initial conditions the first coherent jamming state ($\varphi_o \approx 0$) peaks at 2.5 min which then turns into a stable and smooth, long-term CCW motion ($\varphi_o \approx -\pi/2$) with high correlation[2] from 5 min. Velocity based $\Psi_v$ has an abrupt change at 5 min, while orientation based $\Psi_o$ changes more smoothly.

The results of another typical experiment at higher noise level (*bw* = 1.9 s) can be seen in figure 6. Here correlations are much lower, averaged vector angles tend to have greater noise and ordered states are less stable, they exist only for short periods of time. However, this is the critical noise range where the direction of a coherent moving flock can even change (e.g. from CW to CCW around 8.5 min in figure 6). Changes in the order parameters in this range are more sudden.

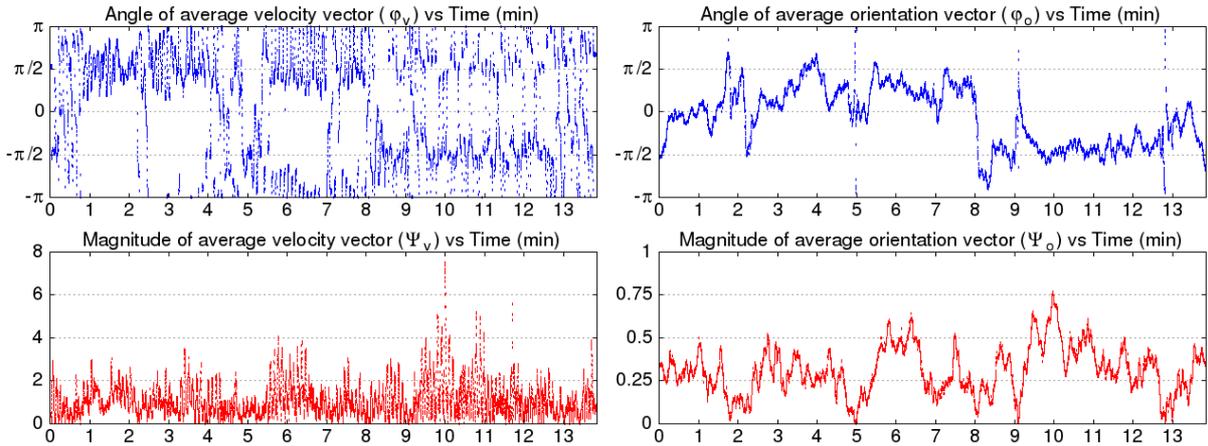

**Figure 6.** Order parameters during an experiment at *bw* = 1.9 s noise level (mid-high range). Starting from random initial conditions there is a tendency of CW motion but without too much coherence. After some small correlation peaks the whole flock changes direction to CCW around 8.5 min. Long term correlated motion can not be observed at high noise levels.

Using 24 normal boats and only 3 leader boats that change direction every 4 minutes between CW and CCW fixed orientations, we can force the whole flock to order in the direction of the leaders in every 4 minute interval with high efficiency. The order parameters (of only the normal boats) of such an experiment can be seen in figure 7.

---

[2] Since the shape of the correlation curves ($C_v$ and $C_o$) usually highly resembles the shape of the average vector magnitude curves ($\Psi_v$ and $\Psi_o$), we shall refer to „high/low correlation" in connection with the high/low range values of both types of order parameters in order to simplify notation.

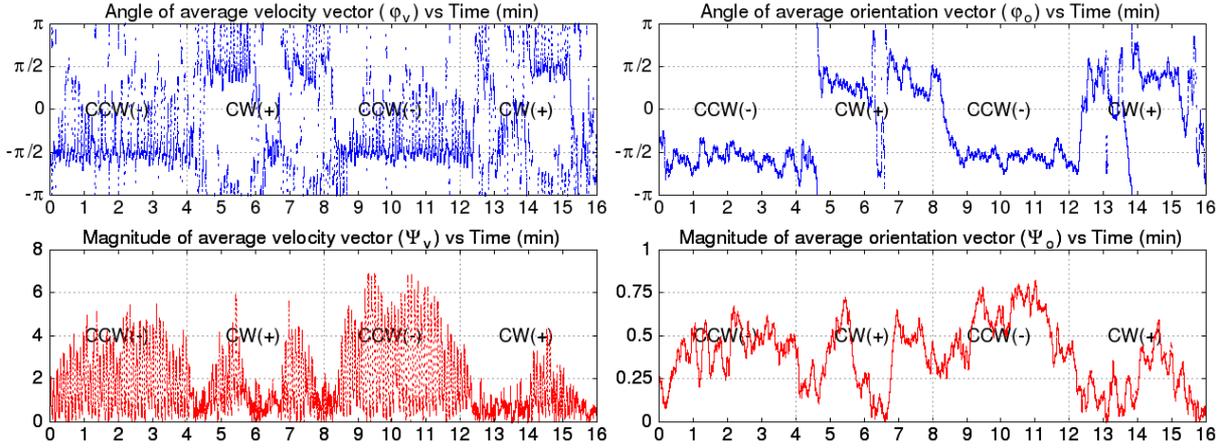

**Figure 7.** Order parameters of the normal boats during an experiment at *bw* = 1.8 s noise level (mid-high range) with 3 leader boats. The fixed direction of the leaders is indicated in the middle of the graphs. The ordering into the same direction as the leaders can be seen clearly from the angle graphs in all 4 parts of the experiment. The correlation values are generally higher than in the normal experiments with the same noise.

The average time needed for achieving an order from random initial conditions was measured in 8-minute experiments either without leader boats or with 3 leader boats with a common CW or CCW direction. We defined the ordered state when the orientations were highly correlated around the ±π/2 angle (CW or CCW) with the following conditions for the order parameters of the normal boats:

$$\left. \begin{array}{r} \dfrac{2}{T}\displaystyle\int_{T/2}^{T} \Psi_o(t)dt > 0.8 \\ \Psi_o(\tau) > 0.8 \\ \left| |\varphi_o(\tau)| - \dfrac{\pi}{2} \right| < \dfrac{\pi}{20} \end{array} \right\}, \qquad (7)$$

where $T$ is the total time of the experiment, $\tau \in [0,T]$ is the time of ordering.

The results are depicted in figure 8. The transition time has large standard deviation but has a definite tendency to decrease with increasing noise level. On the other hand, the coherence of the flock obviously decreases with increasing noise, like in the normal-boat experiments.

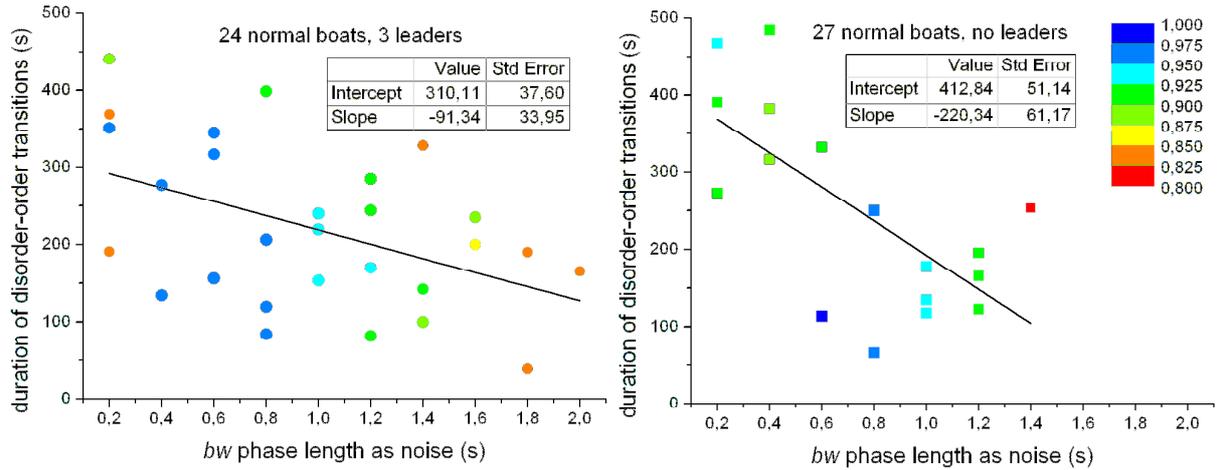

**Figure 8.** The time needed for disorder→order transitions. Each circle is one 8-min experiment with 3 leader and 24 normal boats (left) and 27 normal boats with no leaders (right). Colour indicates average orientation correlation after ordering. Slope and intercept parameters are shown for a linear fit. F-test values of the linear fit (calculated with Origin®, with the null hypothesis that the data cannot be linearly predicted): $F_{left}$=7.24, $p_{left}$=0.01; $F_{right}$=12.97, $p_{right}$=0.003. The transition time has high standard deviation but significantly decreases with increasing noise level in the range where flocking can be observed (negative slope of the fitted lines with low p-values in the F-tests). Leaders extend the noise range where ordering is present and in general slightly reduce the time needed for ordering.

*4.3. State transitions in the noise space*

Plotting the measured PDF of the order parameter values in all the experiments in a common graph as a function of noise level, we can observe interesting state transitions in the noise space as well. Figures 9–11 show the most interesting order parameter PDFs from normal experiments, compared with CW- and CCW-guided leader experiments with 3 leaders.

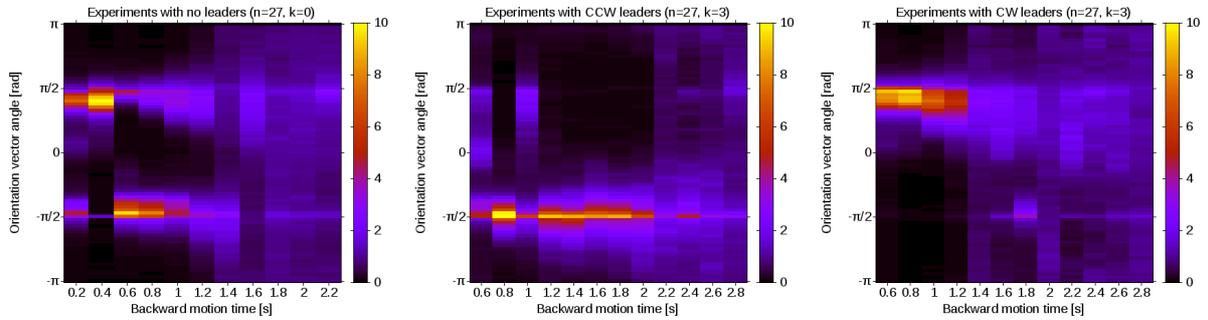

**Figure 9.** PDFs of the orientation vector angles[3] (columns) as a function of noise (horizontal axis). Left: All normal experiments; Middle: all CCW-guided 3-leader experiments; Right: all CW-guided 3-leader experiments. High noise level ($bw > 2$ s) shows disorder on all graphs (close to uniform distribution). Mid noise level shows ordering tendency, but only CW in the CW-guided experiments ($\varphi_v \approx \pi/2$), CCW in the CCW-guided experiments (($\varphi_v \approx -\pi/2$) and both directions in the normal experiments (with symmetric peak probabilities of the orientation angle distribution at $\pi/2$ and $-\pi/2$). Note that the leaders increase the critical noise level between the ordered and disordered states from around 1.4 s to around 2.0 s. Jamming is present occasionally at low noise level, appearing as a small peak at 0 angle, mainly in the non-leader-driven experiments.

---

[3]Since multiple peak probabilities can be present at the same noise level, instead of the order parameter $\varphi_v$ we plotted the full angle probability distribution which gives a smoother result and shows more details.

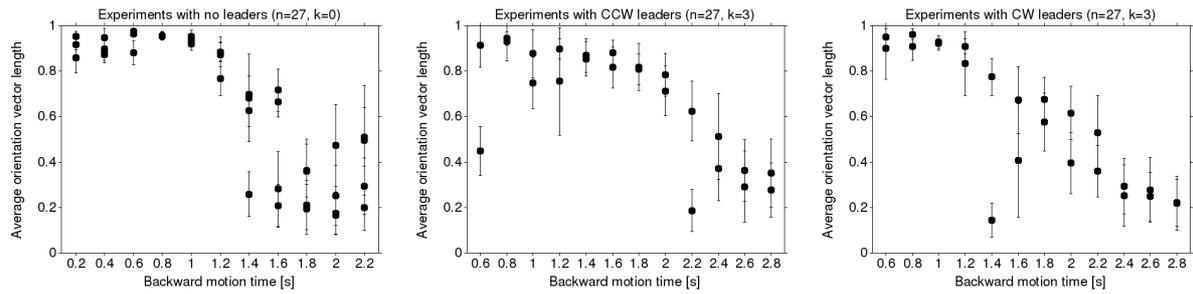

**Figure 10.** Average and standard deviation of $\Psi_o$ (vertical axis) as a function of noise (horizontal axis). Each constant-noise 8-min measurement is plotted separately as one point. Left: All normal experiments; Middle: all CCW-guided 3-leader experiments; Right: all CW-guided 3-leader experiments. All graphs show a nice transition between the low-noise ordered state ($\Psi_o \approx 1$) and a high-noise disordered state ($\Psi_o < 0.3$). In general, the guiding leaders greatly increase coherence in the group in the mid-noise range. At critical noise level all graphs show bi-stability (in the sense that during the 8-minute experiments sometimes the order, sometimes the disorder is dominant), which is the most significant in the no-leader experiments. The order of the disorder-order transition could not be determined from these experiments and this number of boats. The presence of the occasional jamming state on the low noise part of the graphs decreases $\Psi_o$ and increases its standard deviation in some measurements.

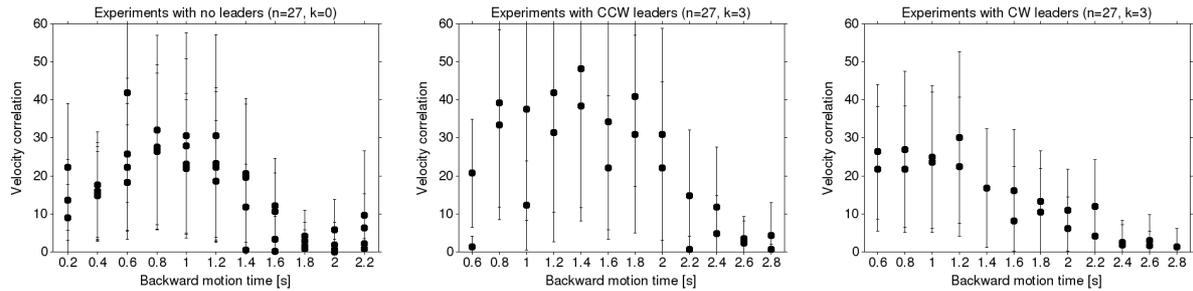

**Figure 11.** Average and standard deviation of $C_v$ (vertical axis) as a function of noise (horizontal axis). Each constant-noise 8-min measurement is plotted separately as one point. Left: All normal experiments; Middle: all CCW-guided 3-leader experiments; Right: all CW-guided 3-leader experiments. The three main collective states can be easily distinguished: (1) disordered motion at high noise level with no velocity correlation; (2) ordered motion at mid noise level with high velocity correlation; (3) mixture of two states (highly ordered motion and jamming) at very low noise level with low velocity correlation again (most significantly in the no-leader experiments). Note that the low velocity correlation at low noise level is generated not only by the jamming phenomena but also by the water current driven by a coherently moving flock (for further details see section 4.5). Current-less results will be provided through simulations in the next chapter. The difference between the CW and CCW graph is due to the slight asymmetry in the boat drive ( see section 4.5).

*4.4. Changing the number of leaders*
The behaviour of the normal boats in the critical noise level (*bw* = 1.8 s) was measured in more detail as a function of the ratio of leader boats. Although the number of leader boats in the experiments was limited to four due to the complexity of these units, even this amount proved to be sufficient for demonstrating their effect on the collective state of the group. The same 8 minute experiments were conducted with 0–4 leaders (with CCW direction preference) to demonstrate that increasing the number of leaders increases the overall order as well. Results can be seen in figure 12.

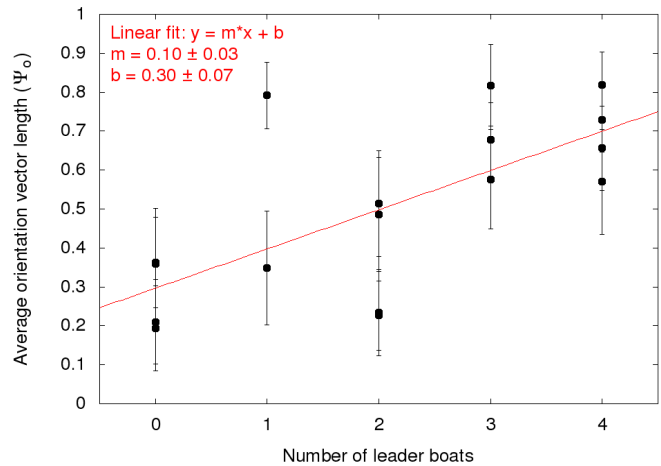

**Figure 12.** $\Psi_o$ and its standard deviation as a function of the number of leader boats at critical noise level ($bw$ = 1.8 s). Each point represents the average of one 8-minute long experiment with fixed leader-boat orientation (CW or CCW). Parameters of a linear regression are shown in the upper left corner. F-test values of the linear fit (calculated with Origin®, with the null hypothesis that the data cannot be linearly predicted): $F$=13.53, $p$=0.002. Order at a given noise level undoubtedly increases with increasing number of leaders (positive slope of the fit with low p-values in the F-test) even though the number of leaders in the experiments is only 5-10% of the whole flock.

*4.5. Error sources*
The most significant source of error in the experiments was the slightly biased CCW tendency of the normal boats, which could not be eliminated and thus affected all results. Nevertheless the aligning tendency due to this biased individual motion could be overcome by the collision-based group behaviour, especially in the critical (mid) noise region between the ordered and disordered states. In that region the leader boats significantly increased coherence of the flock when the leaders were guided in the biased direction and decreased it or changed the flock direction guided in the opposite one.

Another source of error is the state of the battery that affects the speed of the normal boats and also the level of bias in their motion, but this could be minimized by changing the batteries frequently.

Additional unwanted noise in the system comes from the liquid medium in which the boats move. The waves generated by the boats do not affect the motion much and they can be treated as an additional white noise in the orientation and position "updates" of the boats (the amount of which also varies with $bw$). One interesting issue is the net current in the circular pool that can significantly decrease the velocity of the boats in the pool-fixed reference frame when the whole flock is moving in one direction and thus pushing the water in the opposite one. The speed of this current can reach 3-5 cm/s even with our boats of very low speed (10-15 cm/s). Besides modifying the dynamics of the boats to some extent and serving as another error source due to the possibly non-constant velocity profile, the main drawback of this effect is that it reduces the calculated velocity correlations in the ordered state, thereby decreasing the signal-to-noise ratio of our results.

**5. Numerical study of state transitions**
The experimental setup is very limited in terms of flock size and the length of the experiment. To study the effect of flock density and to get a clearer picture of state transitions, an idealized individual-based computational model of the robotic flock was developed. In general, collision-based collective dynamics is strongly influenced by the used model's microscopic rules, which can be diverse in different approaches (cf. [15], [36], [37], furthermore, [38], [13]). Hereby we used a fairly accurate, but in turn computationally intensive model inspired by [39]. The boats were modelled as self-propelled rigid planar objects (figure 13) moving within an annular-shaped area mimicking the pool of the experimental setup. Contact interactions were assumed to be frictionless and perfectly inelastic, which appears to be a reasonable approximation of the experimental setup. For each parameter value, one run of length equivalent of 60 minutes of experiment was conducted.

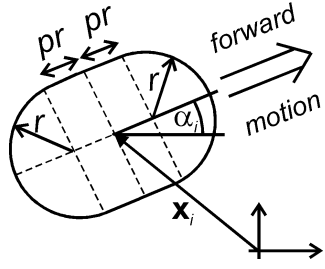

**Figure 13.** Rigid body model of an individual boat. The outline is composed of two semi-circles and two line segments

*5.1. The dynamics of individual boats*
The configuration of boat $i$ is represented by the planar position vector $\mathbf{x}_i$ of its centroid and an orientation angle $\alpha_i$ in a global coordinate system. In the absence of contact with other objects and the boundaries of the region, each boat's position is assumed to evolve according to two differential equations. The first equation

$$\ddot{\mathbf{x}}_i = c_0(v_0 \mathbf{u}(\alpha_i) - \dot{\mathbf{x}}_i) \tag{8}$$

governs the acceleration of the centroid. Dot means derivation according to time; $\mathbf{u}(\alpha_i)$ stands for a unit vector of orientation $\alpha$, i.e. $\mathbf{u}(\alpha_i) = [\cos\alpha_i \; \sin\alpha_i]^T$, where $^T$ is transpose. $v_0$ is the preferred velocity of the boat, which depends on the actual phase (*fw*, *nop* or *bw*). $c_0$ is a damping coefficient determining the rate of convergence to the preferred velocity. The right-hand side of equation (8) can be interpreted as the sum of a propulsive force and linear drag.

The second equation

$$\ddot{\alpha}_i = -c_1 \dot{\alpha}_i^3 + \left(c_2 s_i - c_3 \mathbf{u}^T(\alpha_i)\dot{\mathbf{x}}_i\right)\dot{\alpha}_i + c_4 \eta \tag{9}$$

incorporates variations of the angular velocity due to drag forces on the rudder and the body of the boat. $c_1$, $c_2$ and $c_3$ are positive parameters, and $\eta$ is Gaussian white noise with unit variance. The intelligent boats are actively controlled to achieve a (time-dependent) desired orientation $\beta_i$ corresponding to CW or CCW motion in the pool, via the steering variable $s_i$:

$$s_i = \begin{cases} 1 & \text{if} \quad \pi > \beta_i - \alpha_i > \pi/6, \\ 6(\beta_i - \alpha_i)/\pi & \text{if} \quad |\beta_i - \alpha_i| \leq \pi/6, \\ -1 & \text{if} \quad -\pi < \beta_i - \alpha_i < -\pi/6, \end{cases} \tag{10}$$

while $s_i = 0$ for the rest of the boats. The right-hand side of equation (9) is an idealization of the passive behaviour of the objects: whenever a boat moves forward (i.e. $\mathbf{u}^T(\alpha_i)\dot{\mathbf{x}}_i > 0$), the lateral drag on the backward positioned rudder stabilizes straight motion without active steering ($s_i = 0$). Nevertheless straight backward motion is unstable, and the boats tend to prefer a certain turning rate (specifically $\sqrt{-c_3 \mathbf{u}^T(\alpha_i)\dot{\mathbf{x}}_i / c_1}$) in either direction. Active steering ($s_i \neq 0$) has no effect on a stopped boat, however, it leads to the preference of some nonzero turning rate for moving ones.

Measurements on the experimental setup and the data summarized in table 1 were used to estimate the parameter values of the actual simulation (table 3).

Upon object-object or object-wall contacts, additional contact forces, and instantaneous impulses modify the dynamics of the objects. The outcome of the perfectly inelastic and frictionless interactions is determined by a standard variational method. Details of the simulation method as well as a sketch of the implementation are summarized in an Appendix.

**Table 3.** Parameters of the computational model. The radius of gyration does not show up in equations (8), (9) yet it affects the joint motion of objects in contact.

| Name | Role | Value | Unit |
|---|---|---|---|
| $n$ | number of boats | varied between 20 and 50 | - |
| $k$ | number of leaders | 0 or $0.2 \cdot n$ | - |
| $r$ | size of boat | 5 | cm |
| $p$ | shape of boat | 0.5 | - |
| $\rho$ | radius of gyration of boat | 4.74 | cm |
| $v_0$ | preferred velocity of boat | *fw* phase: 15 | cm/s |
|  |  | *nop* phase: 0 | cm/s |
|  |  | *bw*: phase: –15 | cm/s |
| $c_0$ | convergence rate of velocity | 1 | 1/s |
| $c_1$ | parameters of the dynamics of the angular velocity | 1 | s |
| $c_2$ |  | 0.1 | 1/s |
| $c_3$ |  | 1 | s/cm |
| $c_4$ |  | 0.1 | $1/s^2$ |

*5.2. Simulation results*

Similarly to the experiments, the simulations reveal ordered motion in one (with leaders) or in both directions (without leaders) for low noise levels (figure 14).

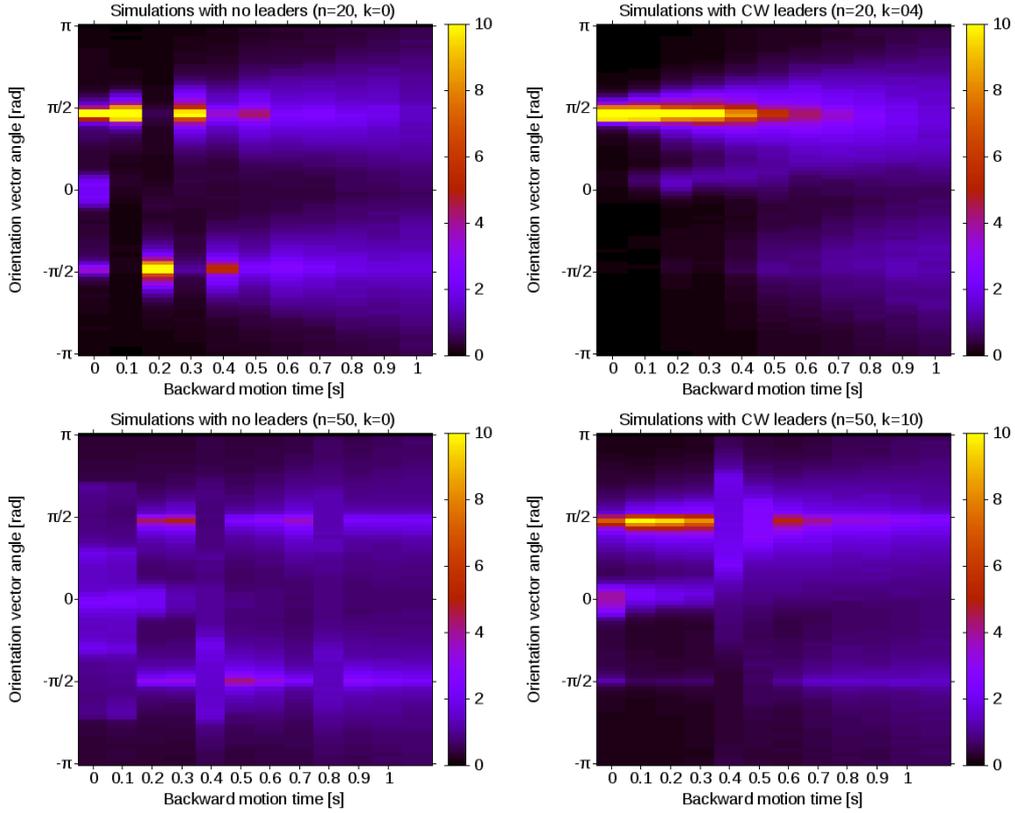

**Figure 14.** PDF of orientation vector angles (vertical axis) as a function of noise (horizontal axis). Top row: $n=20$ objects with $k=0$ (left) and k=0.2$n$ (right) leaders moving CW. Bottom row: same with $n=50$. The vertical asymmetry of the diagrams on the left is due to the fact that only one long simulation was conducted at each noise level and the ordered states are stable, not changing direction frequently. For comparison with corresponding experimental results, see Figure 9.

As the length of the *bw* phase is increased, the degree of ordering drops (figure 15, 16). The transition from order to disorder appears to be at a lower backward motion time (approximately 0.6–0.8*s*) than in the experiments. This difference could be eliminated by tuning the parameters $c_i$ (table 3). The order of the transition is not clear from the simulation results but it seems quite smooth.

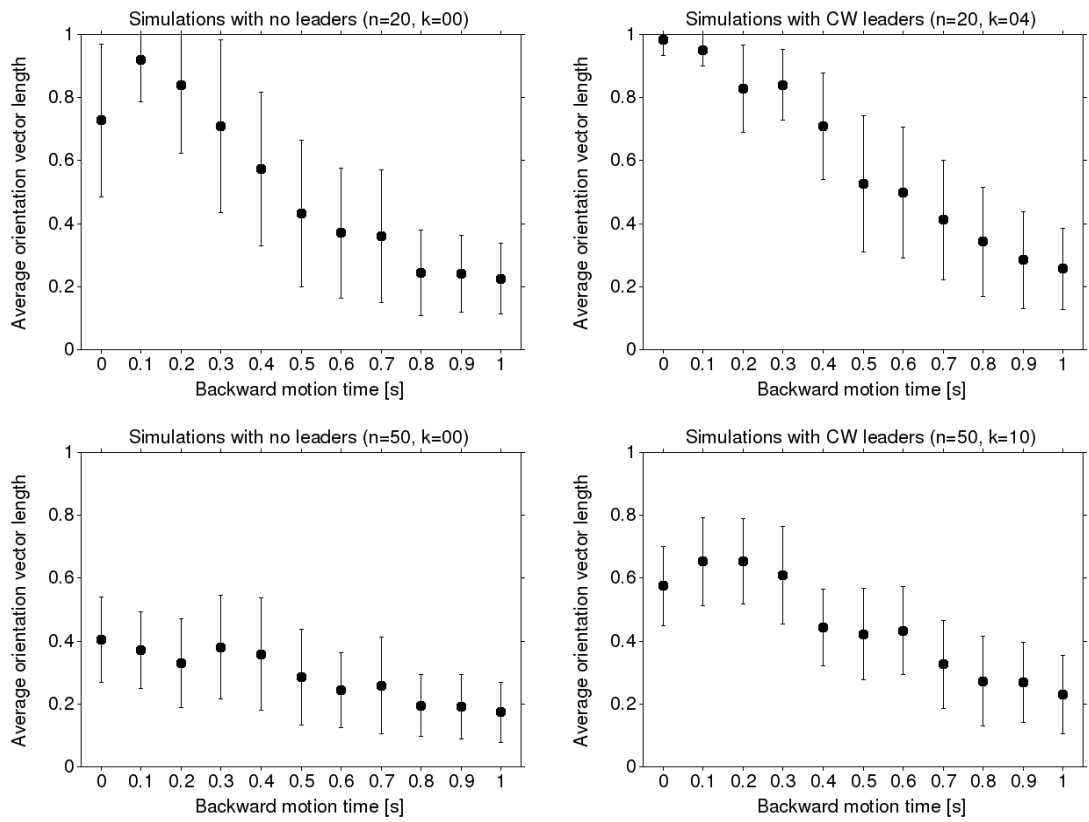

**Figure 15.** Top row: Average and standard deviation of $\Psi_o$ (vertical axis) as a function of noise (horizontal axis) for *n*=20 boats with (right panels) or without (left panels) 0.2*n* leaders. Bottom row: the same with *n*=50. Each point corresponds to one simulation. For comparison with corresponding experimental results, see Figure 10.

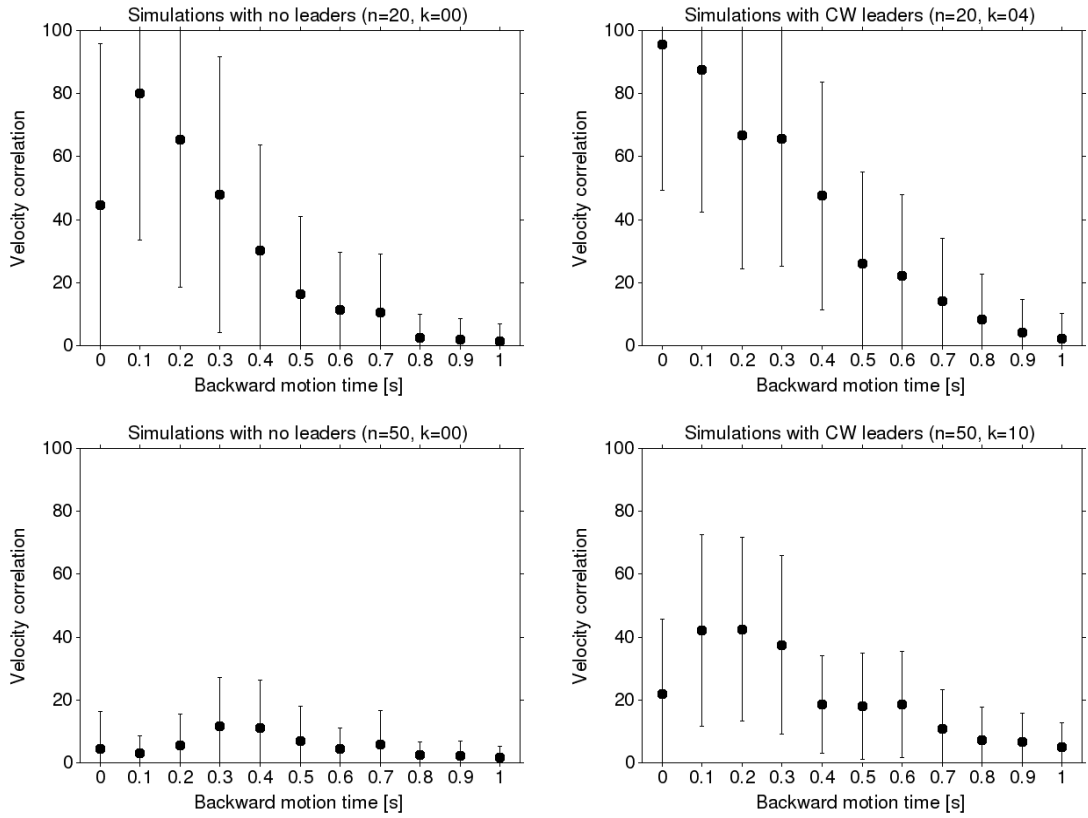

**Figure 16.** Top row: Average and standard deviation of $C_v$ (vertical axis) as a function of noise (horizontal axis) for *n*=20, with or without leaders. Bottom row: same with *n*=50. Each point corresponds to one simulation. For comparison with corresponding experimental results, see Figure 11.

Jamming is observed occasionally in the experimental setup (figure 4) yet it is not frequent (and durable) enough to leave a significant imprint on the PDFs of the order parameters. The same is observed in the simulations if *n*=20. However by increasing the density of the simulated flock, jamming becomes more important. If *n*=50, $C_v$ remains low at low noise levels (figure 16), especially without leaders. This is accompanied by a peak in the distribution of orientation vector angles near zero (figure 14), which is a clear indicator of the jamming state.

To study the jamming – ordered transition, the number of objects was systematically varied between *n*=20 and 50 with backward motion time=0 (figure 17). Both $\varphi_0$ and $\Psi_o$ show a sharp transition between jamming and ordered motion with bistability. The PDFs suggest that the transition from jamming to ordered motion is probably first-order. A solid confirmation of this fact would however require simulations with much bigger flocks.

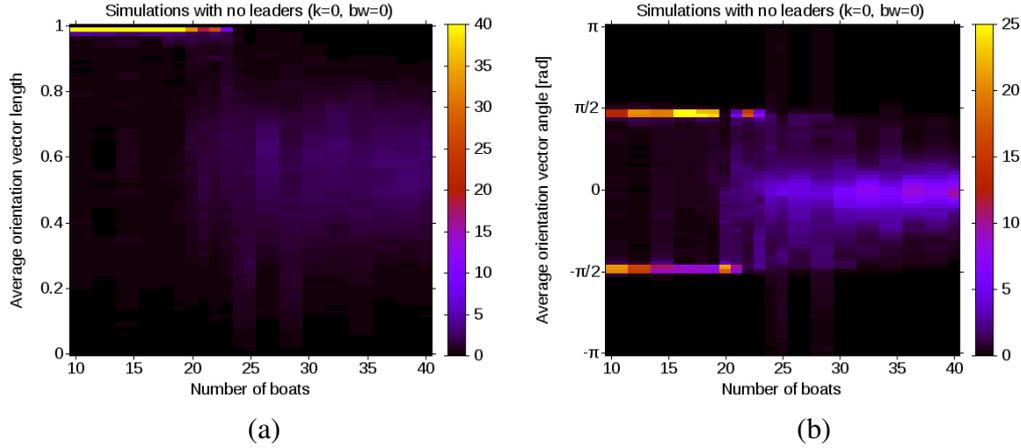

**Figure 17.** PDFs of a) $\Psi_o$ and b) $\varphi_o$ (columns) as a function of density with $k=0$ and $tr=0$. Three runs were conducted for each parameter value, each of length equivalent to 60 minutes of the experiment.

Due to the sensitivity of the jamming behaviour to flock density, there is *negative correlation* between the density and the maximum ordering achievable by tuning the noise parameter. This finding is contrary to the fact confirmed by several works that ordering by contact forces is promoted by sufficiently high density ([8], [13], [40]) in the case of periodic boundary conditions. Yet our finding is in line with the observation that jamming often suppresses other states if the boundary conditions are not periodic ([16] , [17]).

**6. Summary**
We conducted real SPP experiments with radio-controlled boats, communicating through only inelastic collision in an annular pool, and the experiment was accompanied by computer simulation of the system. Our setup had two unique features that provided a step forward from the state-of-the-art. First, we could adjust the noise level of the system. Second, we introduced uniquely driven leader boats to induce specific collective states.

We observed three main states in the collective motion of the boats. The dynamics is dominated by jamming at very low noise level. For long-term ordered motion some noise is needed, while too much noise turns the ordered motion into disordered one. With a constant noise level, the system may switch between states from time to time. Specifically, we observed smooth state transitions from order to disorder and very sudden ones in the case of jamming-ordered transitions. The long-term dynamics is typically dominated by one or two of these states depending on parameters like noise level and flock density. The state transition in parameter space between the ordered and disordered states is smooth in the simulations and is accompanied with bistability at the critical noise level in the experiments. The jamming-order transition is sharp.

The introduction of a small fraction (5-10%) of leader boats has notable effect on the collective dynamics of the system. They control the orientation of the ordered state with high reliability, increase the order at every noise level and, therefore, also increase the critical noise level at which the order-disorder state transition occurs. They are also the most efficient at the critical noise level in pushing the whole system into a dynamic collective state.

With the limited number of boats in our experiment and in the simulations, we cannot precisely determine the order or other properties of the observed state transitions. Nevertheless, we believe that the observed phenomena are the imprints of rich collective dynamics with multiple phases and phase transitions of large non-equilibrium SPP systems.

**Acknowledgments**
Thanks are due to Gergő Somorjai for the great help with the boat-modelling part of the work. This work was supported by the EU ERC COLLMOT project. PV was supported by OTKA grant 72368 and a Bolyai Research Scholarship.

**7. Appendix A**

The appendix is devoted to the presentation of two variational principles, which are used to implement the computational model.

*7.1. Two variational principles*
In the case of object-object or object-wall contacts, the system is subject to nonpenetration constraints. These can be incorporated by adding the appropriate contact forces and impulses as well as their momenta to the right sides of equations (8) and (9). Alternatively, the constrained dynamics can be determined without the explicit calculation of contact forces and impact momenta, using two closely related variational principles:

- According to the Principle of Least Constraint (PLC) [41], the values $\{\ddot{\mathbf{x}}_i, \ddot{\alpha}_i\ i = 1,2,...., n\}$ yielded by equations (8–9) are valid whenever they obey the constraints of the dynamics (i.e. if they do not lead to overlapping objects). Otherwise, the actual accelerations $\{\ddot{\mathbf{x}}_i^{(c)}, \ddot{\alpha}_i^{(c)}\ i = 1,2,...., n\}$ of the constrained system are the minimizers of the quadratic function

$$U_1 = \sum_{i=1,2,...n}(\ddot{\mathbf{x}}_i^{(c)} - \ddot{\mathbf{x}}_i)^2 + \rho^2 \sum_{i=1,2,...n}(\ddot{\alpha}_i^{(c)} - \ddot{\alpha}_i)^2 \qquad (A.1)$$

subject to the constraints of the system; In the formula, $\rho$ stands for the radius of gyration of the objects. Assuming that the objects are of uniform density leads to $\rho = 4.74$ cm.

- Redon et al. [42] proved that in case of a frictionless, plastic impact, the post-impact velocities $\{\dot{\mathbf{x}}_i^+, \dot{\alpha}_i^+\ i = 1,2,...., n\}$ can be obtained from the pre-impact values $\{\dot{\mathbf{x}}_i^-, \dot{\alpha}_i^-\ i = 1,2,...., n\}$ by minimizing the function

$$U_2 = \sum_{i=1,2,...n}(\dot{\mathbf{x}}_i^+ - \dot{\mathbf{x}}_i^-)^2 + \rho^2 \sum_{i=1,2,...n}(\dot{\alpha}_i^+ - \dot{\alpha}_i^-)^2 \qquad (A.2)$$

subject to the non-penetration constraints.

The simulations of this paper are based on the variational approach, which has also been used e.g. in [32].

*7.2. Implementation of the variational principles*
In the simulations, a simple, fixed time-step ODE integrator was combined with quadratic optimization in each time step to simulate the full dynamics of the interacting objects. Due to the analogous forms of the two variational principles, one single optimization scheme implements both of them. Below we summarize the main steps of the integration:

1. At the beginning of the time step, $\mathbf{x}_i(t), \alpha_i(t), \dot{\mathbf{x}}_i(t), \dot{\alpha}_i(t)$ are known, and our goal is to determine them at time $t + \Delta t$.
2. Equations (8,9) are used to obtain $\ddot{\mathbf{x}}_i(t), \ddot{\alpha}_i(t)$, and the following updated values:

$$\left.\begin{aligned}
\dot{\mathbf{x}}_i(t+\Delta t) &\stackrel{def}{=} \dot{\mathbf{x}}_i(t) + \Delta t \cdot \ddot{\mathbf{x}}_i(t), \\
\dot{\alpha}_i(t+\Delta t) &\stackrel{def}{=} \dot{\alpha}_i(t) + \Delta t \cdot \ddot{\alpha}_i(t), \\
\mathbf{x}_i(t+\Delta t) &\stackrel{def}{=} \mathbf{x}_i(t) + \Delta t \cdot \dot{\mathbf{x}}_i(t), \\
\alpha_i(t+\Delta t) &\stackrel{def}{=} \alpha_i(t) + \Delta t \cdot \dot{\alpha}_i(t).
\end{aligned}\right\} \quad (A.3)$$

3. If the new configuration $\mathbf{x}_i(t+\Delta t), \alpha_i(t+\Delta t)$ obeys the constraints (i.e. no objects are overlapping), it is accepted.
4. If not, the non-overlapping constraints are linearized about the point $\mathbf{x}_i(t), \alpha_i(t)$,
5. and the function

$$\sum_{i=1,2,\ldots n}(\mathbf{x}_i^{(c)} - \mathbf{x}_i)^2 + \rho^2 \sum_{i=1,2,\ldots n}(\alpha_i^{(c)} - \alpha_i)^2 \quad (A.4)$$

is minimized, subject to the above linear constraints to obtain the correct updated configuration $\mathbf{x}_i^{(c)}(t+\Delta t), \alpha_i^{(c)}(t+\Delta t)$.
6. The updated velocities are obtained by

$$\begin{aligned}
\dot{\mathbf{x}}_i^{(c)}(t+\Delta t) &= \left[\mathbf{x}_i^{(c)}(t+\Delta t) - \mathbf{x}_i(t)\right]/\Delta t, \\
\dot{\alpha}_i^{(c)}(t+\Delta t) &= \left[\alpha_i^{(c)}(t+\Delta t) - \alpha_i(t)\right]/\Delta t.
\end{aligned} \quad (A.5)$$

Step 5 is a standard quadratic optimization problem, for which many commercial solvers are available. In our simulations, the medium scale quadprog algorithm of Matlab [43] was used. Due to its high computational complexity, this solver can handle up to approximately 100 boats.

**References**
[1] Parrish JK, Hamner WH. Animal Groups in Three Dimensions. Macmillan Publishers Limited. All rights reserved; 1997.
[2] Ballerini M, Cabibbo N, Candelier R, Cavagna A, Cisbani E, Giardina I, et al. Interaction ruling animal collective behavior depends on topological rather than metric distance: evidence from a field study. Proc Natl Acad Sci U S A. 2008;105:1232–1237.
[3] Sumpter DJT, Krause J, James R, Couzin ID, Ward AJW. Consensus decision making by fish. Curr Biol. 2008;18:1773–1777.
[4] Yates CA, Erban R, Escudero C, Couzin ID, Buhl J, Kevrekidis IG, et al. Inherent noise can facilitate coherence in collective swarm motion. Proceedings of the National Academy of Sciences. 2009;106(14):5464–5469.
[5] Szabó B, Szöllösi GJ, Gönci B, Jurányi Z, Selmeczi D, Vicsek T. Phase transition in the collective migration of tissue cells: experiment and model. Phys Rev E. 2006;74:061908.
[6] Vicsek T, Czirók A, Ben-Jacob E, Cohen I, Shochet O. Novel type of phase transition in a system of self-driven particles. Phys Rev Lett. 1995;75:1226–1229.
[7] Huepe C, Aldana M. New tools for characterizing swarming systems: A comparison of minimal models. Physica A: Statistical Mechanics and its Applications. 2008;387(12):2809 – 2822.
[8] Grégoire G, Chaté H. Onset of collective and cohesive motion. Phys Rev Lett. 2004;92:025702.
[9] Aldana M, Dossetti V, Huepe C, Kenkre VM, Larralde H. Phase Transitions in Systems of Self-Propelled Agents and Related Network Models. Phys Rev Lett. 2007 Mar;98(9):095702.

[10]   Aldana M, Larralde H, Vázquez B. Phase transitions in swarming systems: A recent debate. International Journal of Modern Physics B. 2009 Jul;23(18):3459–3483.
[11]   Grégoire G, Chaté H, Tu Y. Moving and staying together without a leader. Physica D. 2003;181:157–170.
[12]   Chaté H, Ginelli F, Montagne R. Simple model for active nematics: quasi-long-range order and giant fluctuations. Phys Rev Lett. 2006;96:180602.
[13]   Grossman D, Aranson IS, Jacob EB. Emergence of agent swarm migration and vortex formation through inelastic collisions. New Journal of Physics. 2008;10(2):023036.
[14]   Blair DL, Neicu T, Kudrolli A. Vortices in vibrated granular rods. Phys Rev E. 2003 Mar;67(3):031303.
[15]   Narayan V, Ramaswamy S, Menon N. Long-Lived Giant Number Fluctuations in a Swarming Granular Nematic. Science. 2007;317(5834):105–108.
[16]   Kudrolli A, Lumay G, Volfson D, Tsimring LS. Swarming and Swirling in Self-Propelled Polar Granular Rods. Phys Rev Lett. 2008 Feb;100(5):058001–.
[17]   Deseigne J, Dauchot O, Chaté H. Collective Motion of Vibrated Polar Disks. Phys Rev Lett. 2010 Aug;105(9):098001.
[18]   SwarmRobot. http://www.swarmrobot.org/;.
[19]   Welsby J, Melhuish C, Lane C, Qy B. Autonomous minimalist following in three dimensions: A study with small-scale dirigibles. In: Proceedings of Towards Intelligent Mobile Robots; 2001. .
[20]   Nardi RD, Holland O, Woods J, Clark A. SwarMav: A Swarm of Miniature Aerial Vehicles. In: Proceedings of the 21st Bristol International UAV Systems Conference; 2006. .
[21]   Conradt L, Roper TJ. Group decision-making in animals. Nature. 2003 Jan;421(6919):155–158.
[22]   Sumpter DJT. Collective Animal Behavior. Princeton: Princeton University Press; 2010.
[23]   Couzin ID, Krause J, Franks NR, Levin SA. Effective leadership and decision-making in animal groups on the move. Nature. 2005;433:513–516.
[24]   Nagy M, Ákos Z, Biro D, Vicsek T. Hierarchical group dynamics in pigeon flocks. Nature. 2010;464:890–893.
[25]   Freeman R, Biro D. Modelling group navigation: dominance and democracy in homing pigeons. Journal of Navigation. 2009;62:33–40.
[26]   Vaughan RT, Sumpter N, Henderson J, Frost A, Cameron S. Experiments in automatic flock control. Robotics and Autonomous Systems. 2000;31(1-2):109–117.
[27]   Correll N, Schwager M, Rus D. Social Control of Herd Animals by Integration of Artificially Controlled Congeners. In: SAB '08: Proceedings of the 10th international conference on Simulation of Adaptive Behavior. Berlin, Heidelberg: Springer-Verlag; 2008. p. 437–446.
[28]   Halloy J, Sempo G, Caprari G, Rivault C, Asadpour M, Tache F, et al. Social Integration of Robots into Groups of Cockroaches to Control Self-Organized Choices. Science. 2007;318(5853):1155–1158.
[29]   Mataric MJ. interaction and Intelligent Behavior. Massachusetts Institute of Technology; 1994.
[30]   Groß R, Bonani M, Mondada F, Dorigo M. Autonomous Self-Assembly in Swarm-Bots. IEEE Transactions on Robotics. 2006;22(6):1115–1130.
[31]   Jasmine. http://www.swarmrobot.org/tiki-index.php;.
[32]   Turgut AE, Çelikkanat H, Gökçe F, ¸Sahin E. Self-organized flocking in mobile robot swarms. Swarm Intelligence. 2008;2:97–120.
[33]   Çelikkanat H, Şahin E. Steering self-organized robot flocks through externally guided individuals. Neural Computing and Applications. 2010 March;.
[34]   Helbing D, Farkas I, Vicsek T. Freezing by Heating in a Driven Mesoscopic System. Phys Rev Lett. 2000;84:1240.
[35]   SwisTrack. http://sourceforge.net/projects/swistrack/;.
[36]   Butt T, Mufti T, Humayun A, Rosenthal P, Khan S, Khan S, et al. Myosin motors drive long range alignment of actin filaments. J Biol Chem. 2010 Feb;285(7):4964–74.
[37]   Sankararaman S, Ramaswamy S. Instabilities and Waves in Thin Films of Living Fluids. Physical Review Letters. 2009 Mar;102(11):118107.

[38] Peruani F, Deutsch A, Bär M. Nonequilibrium clustering of self-propelled rods. Phys Rev E. 2006 Sep;74(3):030904.

[39] Várkonyi PL. Communication and collective consensus making in animal groups via mechanical interactions. J Nonlinear Sci accepted, DOI: 101007/s00332-010-9085-7;.

[40] Mishra S, Baskaran A, Marchetti MC. Fluctuations and pattern formation in self-propelled particles. Phys Rev E. 2010 Jun;81(6):061916.

[41] Baruh H. Analytical Dynamics. WCB McGraw-Hill; 1999.

[42] Redon S, Kheddar A, Coquillart S. Gauss' Least Constraints Principle and Rigid Body Simulations. In: Proceedings of the 2002 IEEE International Conference on Robotics & Automation, Washington DC; 2002. .

[43] http://www.mathworks.cn/access/helpdesk/help/toolbox/optim/ug/quadprog.html;.